\documentclass[journal]{IEEEtran}
\usepackage{etoolbox}
\makeatletter
\patchcmd{\@makecaption}
  {\scshape}
  {}
  {}
  {}
\makeatother
\usepackage{cite}
\usepackage{graphicx}
\usepackage[cmex10]{amsmath}
\usepackage{color}
\begin{document}
\title{Millimeter-Wave Fixed Wireless Access Using IEEE 802.11ay}

\author{Cheng~Chen,
        Oren~Kedem,
        Claudio~R.~C.~M.~da~Silva,
        and~Carlos~Cordeiro% <-this % stops a space
\thanks{Cheng Chen, Claudio R. C. M. da Silva, and Carlos Cordeiro are with Next
Generation and Standards, Intel Corporation, USA. Kedem Oren is with Intel
Communications and Devices Group, Intel Corporation, Israel. Emails:
\{cheng.chen, oren.kedem, claudio.da.silva, carlos.cordeiro\}@intel.com.}}
\maketitle

\begin{abstract}
IEEE 802.11ay defines new PHY and MAC specifications that enable 100
Gbps communications in the 60 GHz millimeter-wave
(mmWave) band. Among the various use cases supported by
IEEE 802.11ay, fixed wireless access, a cost-efficient
high-performance alternative and/or complement to
conventional fixed access, differentiates itself due to its unique
requirements and characteristics. In this article, our goal is to
identify and describe key elements incorporated into IEEE
802.11ay, including scheduling, beamforming, and link
maintenance, that efficiently support fixed wireless access.
IEEE 802.11ay is thus a viable and strong candidate to form the
basis of future generations of standards-compliant (i.e.,
non-proprietary) mmWave fixed wireless access networks.
\end{abstract}


\section{Introduction}
The IEEE 802.11ad amendment ratified in 2012 broke new ground with the introduction of the first multi-gigabit-per-second (multi-Gbps) Wi-Fi technology by using unlicensed spectrum available in the 60 GHz mmWave band. While IEEE 802.11ad supports applications such as instant wireless synchronization, high-speed media file exchange, and wireless cable replacement, requirements of emerging applications go beyond what it can offer. These applications, including augmented reality (AR)/virtual reality (VR),  data center connectivity, vehicle-to-x connectivity, have higher requirements on throughput, reliability, and/or latency. To meet these new requirements, the IEEE 802.11 Task Group ay was formed in 2015 to define PHY and MAC amendments to the IEEE 802.11 standard that enable 100 Gbps communications in the 60 GHz band \cite{11ay}.

Building upon IEEE 802.11ad, IEEE 802.11ay, which had its third draft approved in February 2019, incorporates various new features at the PHY-level, such as channel bonding and aggregation, single-user (SU) and downlink (DL) multi-user (MU) multiple-input and multiple-output (MIMO) transmissions, non-uniform modulation constellation, and orthogonal frequency division multiplexing (OFDM) modulation.
Substantial changes have also been made to the MAC to support these new features.

A few papers have been recently published that study IEEE 802.11ay. For example, \cite{11ay-tutorial} and \cite{11ay-tutorial2} provide an overview of the main design elements of IEEE 802.11ay. A comprehensive description and analysis of channel models and the single-carrier (SC) PHY are presented in \cite{11ay-ChannelModel} and \cite{11ay-phy}, respectively. Two important beamforming training procedures, beam refinement protocol (BRP) transmit sector sweep (TXSS) and asymmetric beamforming training, are reviewed in \cite{11ay-bf} with detailed discussions, whereas the authors in \cite{11ay-mimo} focus on enhanced MU-MIMO beamforming protocol.

One of the usage models supported by IEEE 802.11ay is mmWave fixed wireless access, also referred to as mmWave distribution network\footnote{Since mmWave fixed wireless access is also referred to as mmWave distribution network in IEEE 802.11ay, we use these two terms interchangeably throughout this paper.}\cite{TDD-UseCase}. This use case can be used to enable different deployment scenarios, such as broadband residential access, Wi-Fi access point (AP) and small cell backhaul, and home media sharing. It is seen as a promising cost-efficient high-performance alternative and/or complement to conventional fixed access networks, which offers many advantages to service providers, including fast time-to-market through rollout speedup, high coverage, low upfront cost, and less coordination with various building owners. Fig. 1 depicts an areal view of mmWave distribution network.

Using mmWave communications as the backhaul for fixed wireless access has been studied for a long time \cite{FWA-1, FWA-2, FWA-3}. However, most of such implementations rely on proprietary technology using licensed spectrum \cite{FWA-4, FWA-5}. Existing mmWave standards in unlicensed spectrum, such as IEEE 802.11ad and IEEE 802.15, do not incorporate necessary protocols to support fixed wireless access. IEEE 802.11ay is the first mmWave standard to exclusively define a complete set of PHY and MAC technologies to enable fixed wireless access in unlicensed 60 GHz spectrum.

Using standardized technology lowers implementation cost and supports interoperability among multiple vendors. By operating in the mmWave band, IEEE 802.11ay provides much increased capacity compared to other Wi-Fi systems that operate in microwave bands. Additionally, since mmWave links are highly directional, it also presents significant opportunities for spatial reuse, offering more flexibility in the deployment of fixed wireless access systems. Actually, the value of fixed wireless access using 60 GHz band has already been shown, for example, in the Terragraph system prototype introduced by Facebook \cite{Terragraph}.

In order to accommodate this new use case, a number of features were incorporated into IEEE 802.11ay that define techniques and mechanisms necessary to support mmWave distribution networks in 60 GHz band. This paper studies the primary features of mmWave distribution networks and the corresponding design elements IEEE 802.11ay has adopted in terms of the AP-STA communication model.

The remainder of this article is organized as follows. In Section \ref{Sec:mmWave_Distribution_Network_in_80211ay}, we provide an overview of the main requirements of mmWave distribution networks and identify design challenges for IEEE 802.11ay. We then present how IEEE 802.11ay addresses these challenges related to scheduling, beamforming, and link maintenance in Section \ref{Sec:Scheduling}, Section \ref{Sec:Beamforming}, and Section \ref{Sec:Link Maintenance}, respectively. We conclude in Section \ref{Sec:Conclusions}.

\begin{figure}[htbp]
\centering
\includegraphics[width=0.5\textwidth,height=0.3\textwidth]{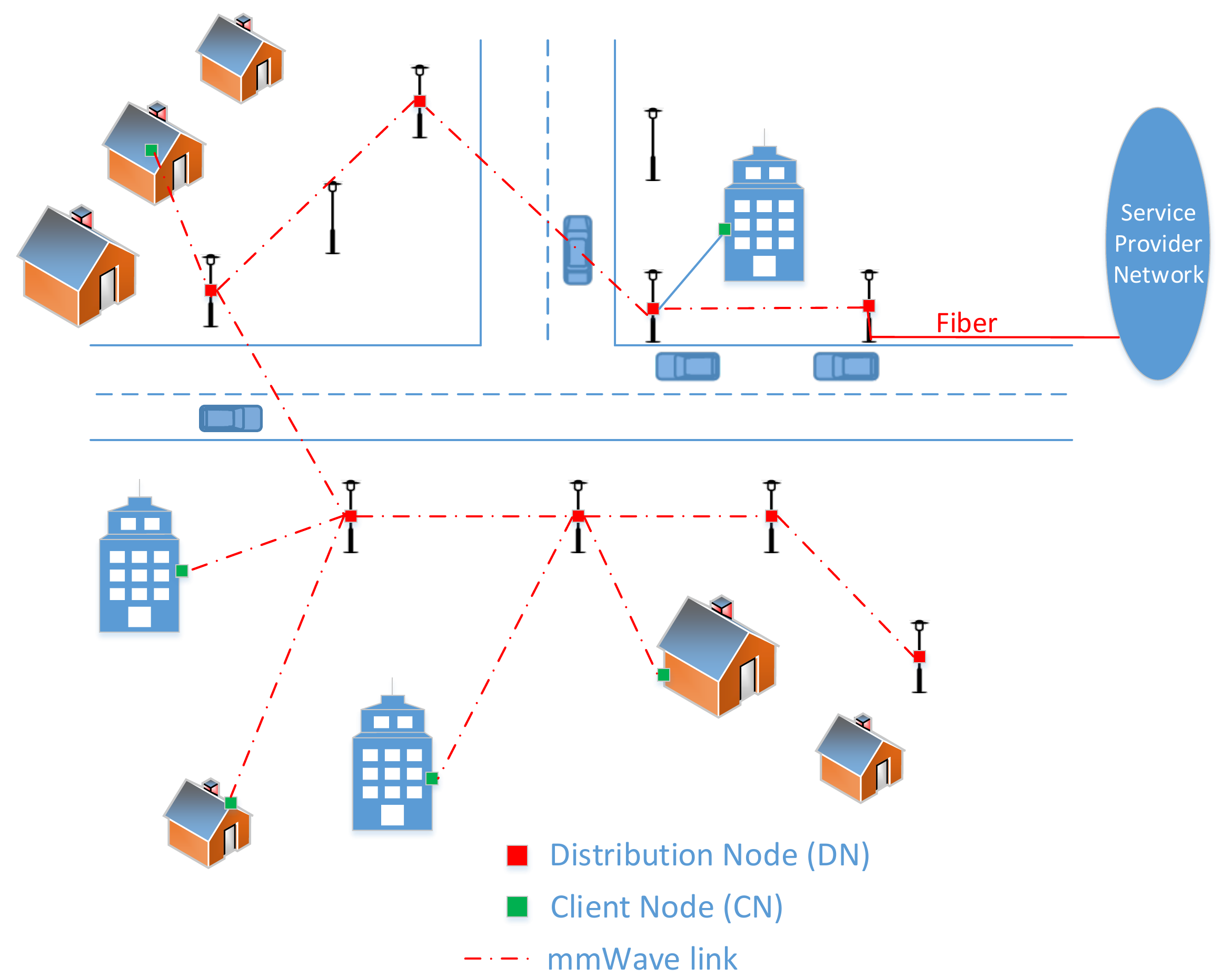}
\caption{Areal view of mmWave distribution network.}
\label{Fig:TDD Use Case}
\end{figure}

\section{mmWave Distribution Network in IEEE 802.11ay}\label{Sec:mmWave_Distribution_Network_in_80211ay}
In this section, we describe the main features and requirements of mmWave distribution networks, as well as the challenges in terms of scheduling, beamforming, and link maintenance.

The mmWave distribution network model considered by IEEE 802.11ay has a multi-hop multi-point mesh network topology with a relatively large density of distribution nodes (DNs) and client nodes (CNs). A centralized controller, which typically stays in the cloud, is designated for network management. A DN is composed of multiple sectors capable of communication in different directions, each of which managed by an AP. A CN is typically a station (STA) served by a nearby DN. DN-CN communication uses the AP-STA communication model defined in IEEE 802.11. For DN-DN communication, one DN takes the role of STA and hence the AP-STA model still applies. DNs can be placed at street poles (e.g., lamppost), and serve individual homes, or serve office buildings, as shown in Fig. 1.

Devices in mmWave distribution networks operate in outdoor environment with line-of-sight (LOS) channels under most conditions. The distance per hop between the paired DNs is less than 300 meters. The downlink sustainable data rate requirement is more than 4 Gbps per DN, and the latency requirement is less than 15 ms \cite{TDD-UseCase}.

\subsection{Scheduling}
The strict requirement of high data rate and low latency for mmWave distribution networks calls for careful interference management and brings significant challenges to the scheduling protocol in IEEE 802.11ay, which is built on the hybrid MAC approach defined in IEEE 802.11ad that supports both contention-based medium access and scheduled channel time allocation, corresponding to contention-based access periods (CBAPs) and service periods (SPs), respectively. In CBAPs multiple stations (STAs) contend for transmission based on the 802.11 enhanced distributed coordination function (EDCF). The random nature of co-channel interference in CBAPs leads to issues such as exposed and hidden node problems, and therefore prevents any meaningful interference management. For SPs, even though transmissions are scheduled, the network still suffers from uncontrolled interference as SPs of different APs are not coordinated and can overlap in time.

In order to minimize interference, mmWave distribution networks require coordinated, scheduled, and contention-free allocations across all links, which are decided by a centralized controller that manages transmissions in the entire network. Moreover, downlink and uplink transmissions have to be strictly separated to ease the scheduling complexity and further mitigate interference. Since neither CBAPs nor existing SPs are capable of meeting these requirements, IEEE 802.11ay creates a new time division duplex (TDD) scheduling protocol\footnote{The naming convention of IEEE 802.11ay for all mechanisms related to mmWave distribution networks is to append TDD at the beginning. For example, TDD scheduling, TDD beamforming, and TDD link maintenance.}, which is presented in Section \ref{Sec:Scheduling}.

\subsection{Beamforming}
Beamforming is defined as the procedure used to determine appropriate receive and transmit antenna configurations for a pair of STAs to communicate. The beamforming initiator refers to the STA that triggers the beamforming procedure while the responder is the targeted STA with which the initiator plans to perform training. IEEE 802.11ad defines a complete flow of beamforming, consisting of an initial coarse-grained sector-level sweep (SLS) and a subsequent beam refinement phase (BRP). IEEE 802.11ay further introduces various enhancements which not only improve beamforming performance, but also extend it to support new applications and transmission modes \cite{11ay-bf}.

In mmWave distribution networks, beamforming is generally triggered after a newly installed node gets registered via an off-channel mechanism, and then boots up and continuously sweeps beams in receive mode in order to establish a link with a mmWave AP within a nearby DN. 
%The new node first gets registered via an off-channel mechanism, for example, registration at a website, so that the centralized controller is aware of the fact that the new node is joining the network.

Existing beamforming protocols assume STAs employ a quasi-omni antenna configuration for initial beamforming training. However, the difference between the gain of directional and quasi-omni antenna arrays is much higher than the link budget between the lowest modulation and coding scheme (MCS 0) and higher data MCSs. Consequently, the transmission range while transferring data when both sides use directional antenna configurations is much larger than the transmission range when one side of the link uses quasi-omni and the other uses directional antenna configuration. Considering this, directional beams are always used in mmWave distribution networks with antenna reciprocity, which means the transmit and receive antenna configurations of a given STA are the same.

Another challenge comes with the fact that the downlink and uplink transmissions in mmWave distribution networks are separated, while existing beamforming protocols require bidirectional traffic to enable instant exchange of frames between the initiator and responder.

Considering the complexity of modifying existing beamforming flow to support mmWave distribution networks, IEEE 802.11ay defines a separate beamforming protocol for this use case, which is described in Section \ref{Sec:Beamforming}.

\subsection{Link Management and Maintenance}
Link maintenance is an essential part for Wi-Fi communications as the wireless channel conditions can change dramatically due to various effects such as fading, blockage, and interference. It is particularly important for 60 GHz transmissions which are highly directional.

Compared to other use cases, mmWave distribution networks introduce several new messages for link maintenance. There are specific requirements for global time synchronization to drive transmission and reception of frames using negotiated time slots. Otherwise, neighboring links may  potentially interfere with each other. Moreover, transmit power control (TPC) is critical in terms of link management and interference mitigation across the network.

Section \ref{Sec:Link Maintenance} introduces how IEEE 802.11ay modifies existing link maintenance protocols to accommodate these requirements.

\section{TDD Scheduling}\label{Sec:Scheduling}
The scheduling mechanism developed for mmWave distribution networks in IEEE 802.11ay is built upon the concept of scheduled SPs defined in IEEE 802.11ad, where an SP is a contention-free period assigned to the communication between a dedicated pair or group of nodes.

A new type of SP allocation, TDD SP, is defined to be dedicated to the transmission and reception between an AP and its associated STAs within mmWave distribution networks. Only STAs that support TDD channel are allowed to access TDD SPs. STAs that do not support this mode regard TDD SPs as conventional SPs not assigned to them, and therefore will not attempt to access the channel. The main features of TDD scheduling include the following:

\begin{itemize}
\item 1)
\textbf{Slotting structure of time allocation}: TDD scheduling follows a strict slotted structure, where a period of time allocation is divided into smaller chunks of slots, each of which is then assigned to transmissions between (a) dedicated pair(s) of nodes.
\item 2)
\textbf{Dynamic scheduling assignment}: The assignment of slots to STAs is highly dynamic. It is determined by the centralized controller so that coordinated time allocations across different APs and STAs are possible.
\item 3)
\textbf{Unidirectional traffic within a slot}: Within any given slot, only unidirectional transmission is allowed. Consequently, a STA either transmits or receives during a slot, but is not allowed to do both.
\end{itemize}

Considering time allocation, slot structure, and access assignment of TDD SPs change on different time scales, as indicated in Table \ref{Table:TDD_Scheduling}, the overall scheduling protocol is performed in three levels, as illustrated in Fig. 2,

a) The AP uses an Extended Schedule element to schedule TDD SPs for STAs operating with TDD channel access.

b) Slot structure information of the allocated TDD SP is defined in TDD Slot Structure element.

c) The access assignment of STAs to the allocated TDD SP is defined with a TDD Slot Schedule element.

\subsection{TDD SP Allocation}
A TDD SP is scheduled by the Extended Schedule element defined in IEEE 802.11ad. A unique indication bit is added to differentiate it from a conventional SP. The Extended schedule element is included in broadcast messages like directional multi-gigabit (DMG) Beacon frame. This allows other STAs to be aware of SP allocations scheduled for TDD operations, hence improving coexistence.

\subsection{TDD SP Slot Structure}
Each TDD SP consists of one or more consecutive and adjacent TDD intervals, and each TDD interval comprises one or more TDD slots. The parameters of the TDD structure, including number of TDD slots per TDD interval, the start time and duration of each TDD slot, are contained in the TDD Slot Structure element.

An AP advertises the TDD Slot Structure element to each STA that is expected to transmit or receive during a TDD SP. Once received, a STA adopts the TDD structure within the element for all TDD SPs identified by the same Allocation ID subfield value, and the structure remains in effect until the STA receives an updated TDD Slot Structure element. Nevertheless, slot structure is not likely to change frequently in an active network.

\subsection{TDD SP Access Assignment}
The access protocol of TDD slots for the STAs is strictly scheduled, which means a STA is only allowed to communicate with an AP in specific slots where it is assigned. Moreover, in each assigned TDD slot, only unidirectional traffic is allowed, and different types of traffic will be given different priorities depending on the category of the TDD Slot. All these aspects are described by the TDD Slot Schedule element.

\begin{figure}[htbp]
\centering
\includegraphics[width=0.5\textwidth,height=0.3\textwidth]{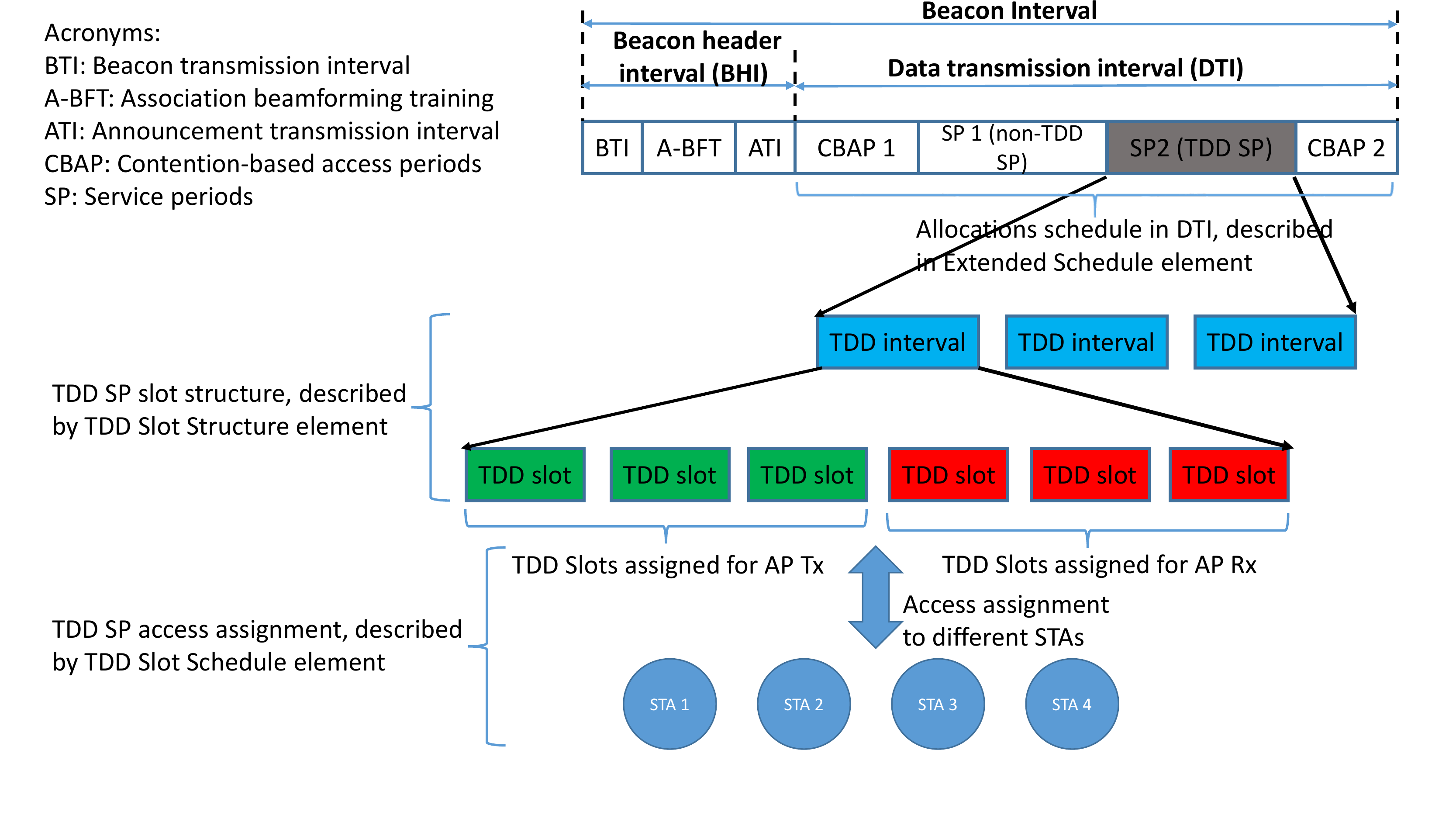}
\caption{The scheduling structure of a TDD SP.}
\label{Fig:TDD_Scheduling}
\end{figure}

\begin{table}[htbp]
\centering
\caption{Typical time scale of different time periods.}
\begin{tabular}{|c|c|c|c|c|}
\hline
Time period &TDD Slot &TDD Interval &TDD SP & BI \\
\hline
Typical duration &66$\mu$s &1.6ms & 25.6ms & 300ms\\
\hline
\end{tabular}
\label{Table:TDD_Scheduling}
\end{table}

%Specifically, the Bitmap and Access Type Schedule field within TDD Slot Schedule element uses pairs of consecutive two bits to define the type of each TDD slot, as well as the access permission of a STA to these TDD slots covered by the entire bitmap. Depending on the value of the encoding, each TDD slot can be unassigned, assigned as transmit, assigned as receive, or unavailable, i.e., temporarily unassigned but reserved for future use. In addition, the Slot Category Schedule field defines the corresponding traffic priority. For example, in a Basic TDD slot, the transmission of all frame types are allowed, but Control and Management frames are prioritized. In contrast, in a Data-only TDD slot, only Data frames and BlockAckReq frames are allowed.

\subsection{Delayed Acknowledgement} 
The nature of unidirectional traffic in a TDD slot causes a
problem with the acknowledgement process, where the receiver is expected to send an Ack or BlockAck frame to the transmitter immediately after a successful transmission. In order to resolve this issue, when operating in a TDD SP, a STA does not acknowledge the reception immediately in the same TDD slot. Instead, it transmits the Ack or BlockAck frame at the start of the earliest Basic TDD slot in which the STA is assigned to transmit. Since control frames are given priority in a Basic TDD slot, it protects the delivery of the acknowledgement from possible scheduling conflicts.

There are several other optional frame exchanges that solicit immediate responses, such as RTS/DMG CTS, Grant/Grant Ack, etc. Unlike acknowledgement frames, delaying the responses to later TDD slots does not help with the intended functions of these frame exchanges. As a result, the usage of these control or management frames is prohibited within a TDD slot.

\begin{figure*}[htbp]
\centering
\includegraphics[width=1.0\textwidth,height=0.45\textwidth]{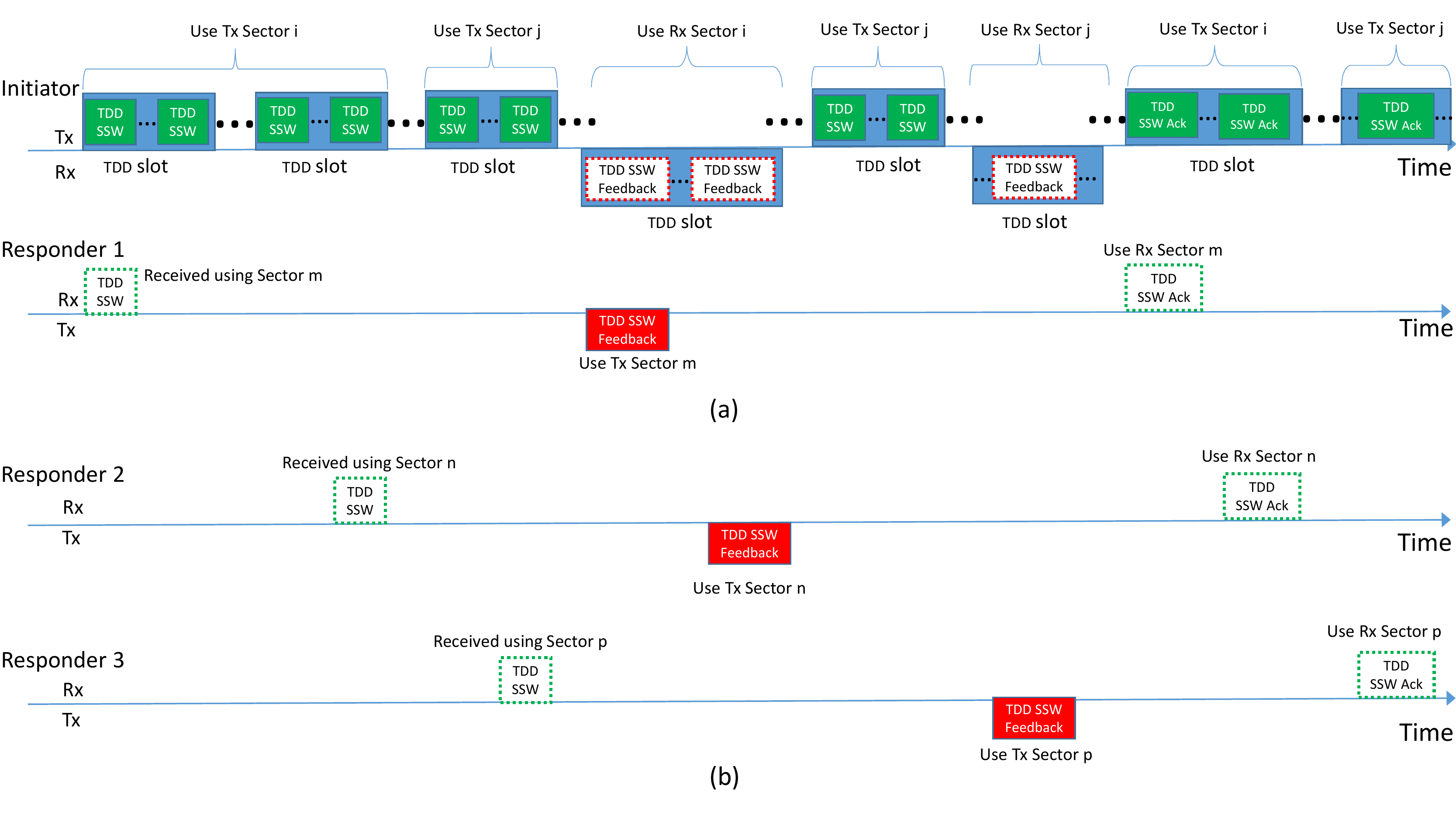}
\caption{Examples of TDD beamforming: (a) TDD individual beamforming, (a) + (b) TDD group beamforming.}
\label{Fig:TDD_Beamforming}
\end{figure*}

\section{TDD Beamforming}\label{Sec:Beamforming}
IEEE 802.11ay defines a dedicated TDD beamforming protocol exclusively for STAs operating in mmWave distribution networks, which includes three modes:

\begin{itemize}
\item
\textbf{TDD individual beamforming}: An initiator transmits a series of TDD Sector Sweep (SSW) frames and a single responder performs receive sector sweep. Through exchanging additional frames, both the initiator and responder identify one or more combinations of transmit beam on the initiator side and receive beam on the responder side that enable communication between the two STAs.
\item
\textbf{TDD group beamforming}: This is similar to TDD individual beamforming except that the procedure is extended to include multiple responders.
\item
\textbf{TDD beam measurement}: An initiator transmits a series of TDD SSW frames while one or more responders sweep their receive sectors to take measurements. The primary difference with TDD individual/group beamforming is here responders do not transmit any frames to the initiator in the PHY/MAC layer. Instead, responders report the measurement results to the centralized controller using high-layer management entity, which helps to keep track of interference within the network.
\end{itemize}

\subsection{Overall Beamforming Process}
TDD beamforming protocol is performed in a TDD SP and assumes antenna reciprocity for both initiator and responder(s) STAs. The general procedures are as follows.
\begin{itemize}
\item 1)
The initiator sends multiple TDD SSW frames using each of its transmit sectors. Each TDD SSW frame includes the following information:
\begin{itemize}
\item
The index of the transmit sector used to send the TDD SSW frame.
\item
The time offsets for each responder that receives the TDD SSW frame should a) send a TDD SSW Feedback frame back to the initiator as response, and b) should expect to receive a TDD Ack frame from the initiator. In case of TDD group beamforming, these two time offsets can be different across responders.
\end{itemize}
\item 2)
If a responder receives a TDD SSW frame, it  sends a TDD SSW Feedback frame at the time indicated in the decoded TDD SSW frame. After the transmission, it should be ready to receive the TDD SSW Ack frame from the initiator, at the time also included in the decoded TDD SSW frame.
\item 3)
Once a TDD SSW Feedback frame is received, the initiator transmits a TDD SSW Ack frame as an acknowledgement. The beamforming flow ends when the initiator sets the End of Training subfield to one in TDD SSW or TDD SSW Ack frame.
\end{itemize}

After the completion of a TDD beamforming procedure, a link is established between the initiator and responder. They can therefore perform association and authentication, exchange capabilities and network configuration, and start data transmissions.

Examples of TDD individual and group beamforming are illustrated in Fig. \ref{Fig:TDD_Beamforming}.

\subsection{Initiator Operation}
During TDD individual or group beamforming, the initiator sends multiple TDD SSW frames through each of its transmit sectors. In each TDD SSW frame, the initiator includes timing information for the responder that receives the TDD SSW frame to transmit TDD SSW Feedback frame. Note that the initiator does not know whether a TDD SSW frame is received by a responder or not, so it always switches to receive mode at the indicated time, and sets its receive antenna to the same sector used to send the TDD SSW frame. If it receives a TDD SSW Feedback frame, it then transmits a TDD SSW Ack frame to acknowledge the receipt at the time indicated in the corresponding TDD SSW frame. In the TDD SSW Ack frame, the initiator includes information of the sector used to transmit the TDD SSW Ack frame, the sector used by the responder to transmit the TDD SSW Feedback frame, as well as the measured signal-to-noise ratio (SNR) of the decoded TDD SSW Feedback frame. Additionally, it includes the time offsets to exchange Announce frames containing STA capabilities and network configuration. The initiator indicates the termination of beamforming by setting the End of Training field to one in either TDD SSW or TDD SSW Ack frame. %This may happen when the initiator and the responder have found a transmit and receive antenna configuration that is capable of data communication.

\subsection{Responder Operation}
During TDD individual or group beamforming, the responder first sweeps its receive antenna through all of its receive sectors and dwell on each sector for a specific time. When it receives a TDD SSW frame, it determines the timing information to send TDD SSW Feedback frame to the initiator, as well as when to expect a TDD SSW Ack frame from the initiator. When it comes time to send the TDD SSW Feedback frame, the responder uses the sector from which it received the corresponding TDD SSW frame with the best link quality. When it comes time to receive the TDD SSW Ack frame, the responder sets its receive antenna to the same sector that was used to transmit the TDD SSW Feedback frame. If the TDD SSW Ack frame includes a positive indication for End of Training field, the responder will use this sector in all subsequent transmissions until the next beamforming procedure is initiated.

\subsection{TDD Beam Measurement}
The flow of TDD Beam Measurement is nearly the same with TDD individual or group beamforming. The differences for the behaviors of the initiator and responder are as follows. For the initiator, the transmitted TDD SSW frames do not include the two offsets, i.e., the offset for the responder to transmit the TDD SSW Feedback frame and the offset for the responder to receive the TDD SSW Ack frame. Instead, a TDD slot countdown is presented to indicate when the transmission of TDD SSW frames ends. With this change, the responder does not transmit any frames to the initiator, and the initiator does not send TDD SSW Ack frame.

\section{TDD Link Management and Maintenance} \label{Sec:Link Maintenance}
IEEE 802.11ay addresses link management and maintenance for mmWave distribution networks mainly by reusing the existing link maintenance protocol, with some necessary modifications and enhancements.

\subsection{Link Maintenance}
Similar to other use cases, link management in mmWave distribution network is achieved by requesting and reporting relevant network information, and several management messages are introduced to facilitate the process. The three main messages are Heartbeat, Keep Alive, and Uplink Bandwidth Request.

Heartbeat message is used to update parameters and schedule transmit/receive slots for associated STAs. Keep Alive message is transmitted periodically to track the existence of a communication link between peers and indicate the AP is able to communicate, as well as to negotiate receive time slots with the peer STA. Uplink bandwidth request message is used to report the aggregate state of traffic and request bandwidth grant from the STA to AP.

Instead of defining three messages separately, IEEE 802.11ay reuses existing Announce frame and includes specific information elements to enable the functions. This method is simple, efficient and minimizes overhead. It also provides extensive flexibility thanks to the following characteristics of Announce frames: an Announce frame can either require acknowledgement or not. Its transmission can be either unicast or broadcast, either encrypted or non-encrypted. Furthermore, Announce frame can carry different information elements to serve various functions.

The primary elements incorporated into IEEE 802.11ay to support TDD link maintenance are: First, several optional fields are added to DMG Link Margin element, an existing information element used for link adaptation, to provide additional information such as parameters across receive chains, protocol data units (PPDUs), low-density parity-check (LDPC) codewords, single-carrier blocks, OFDM symbols, and transmit power control. Second, a new information element, TDD Bandwidth Request element, is defined to address bandwidth reservation request, which includes relevant information such as queue size, traffic arrival rate, and traffic identifier.

\subsection{Time Synchronization}
Relying on centralized scheduling, mmWave distribution networks maintain global time synchronization to avoid network self-interference. Therefore, link maintenance is easier if at least one of the two connecting nodes is in global sync mode. IEEE 802.11ay defines a new TDD Synchronization element to indicate the quality of the clock at a STA available to its peers. However, DMG STAs that support TDD mode are free to use different protocols for time synchronization.

\subsection{Periodic Link Measurement Report}
Another important feature of mmWave distribution networks is frequent and periodic link measurement. Unfortunately, previous link measurement protocol only supports on-demand link measurement which requires the use of the paired Link Measurement Request and Link Measurement Report frame exchange. However, the requirement of transmitting Link Measurement Request frame increases overhead for periodic reporting. Furthermore, it is susceptible to the loss of Link Measurement Request frame. Considering this, IEEE 802.11ay extends the link measurement protocol to accommodate unsolicited periodic link measurement reporting.

\begin{figure}[htbp]
\centering
\includegraphics[width=0.5\textwidth,height=0.3\textwidth]{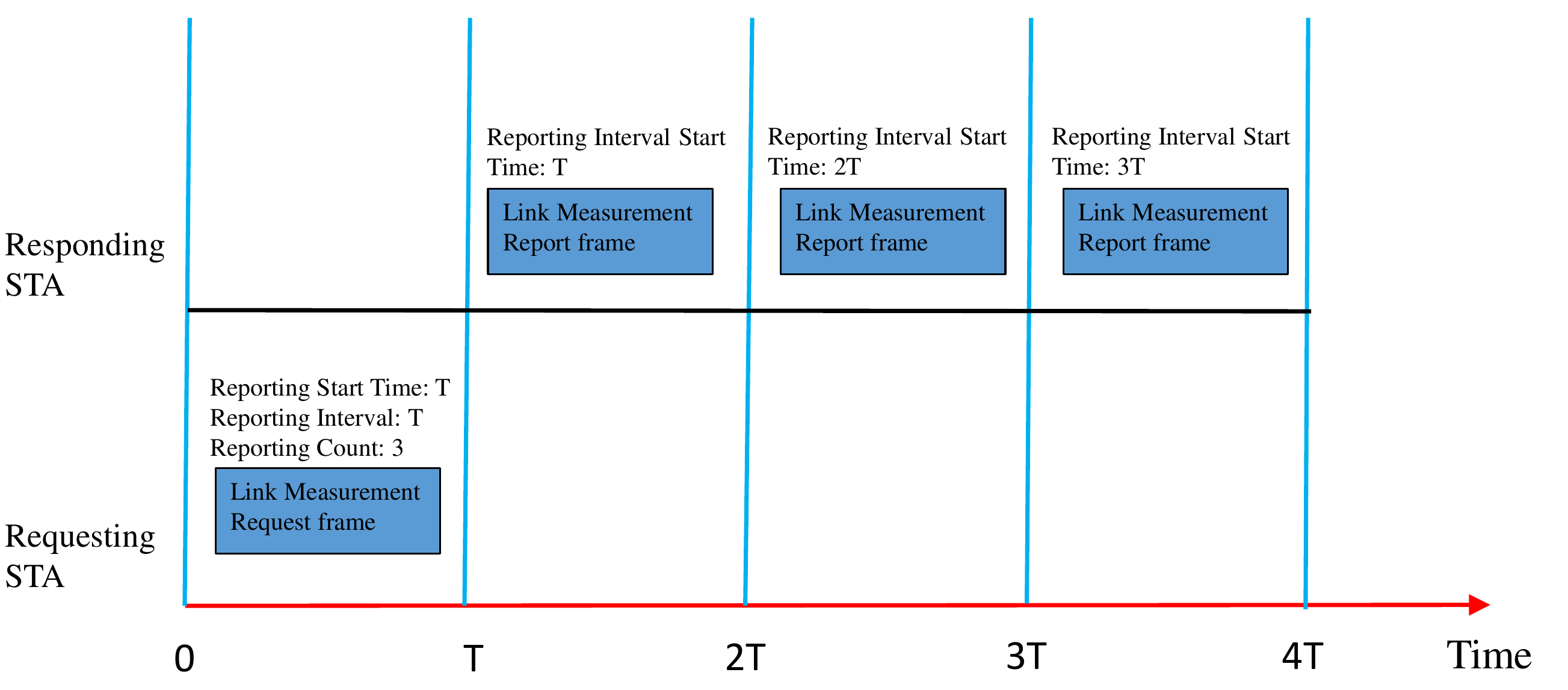}
\caption{An example of periodic link measurement report.}
\label{Fig:TDD_Link_Maintenance}
\end{figure}

Specifically, a new information subelement called Periodic Report Request is defined, which specifies information such as report start time, report interval and report count. When received, a responder can either accept or reject the periodic report request. If accepted, the responder will send Link Measurement Report frames periodically based on the agreed schedule, including information like received channel power indicator (RCPI) and received signal to noise indicator (RSNI). Fig. \ref{Fig:TDD_Link_Maintenance} shows an example of this mechanism.

\subsection{Transmit Power Control}
Transmit power control is one of the techniques mmWave distribution networks utilize to address interference mitigation. IEEE 802.11ay introduces an enhanced TPC protocol that incorporates a series of new features including extension of TPC to multiple transmit and receive chains, support for TPC based on periodic link measurement, and more information parameters. These optional fields are appended to DMG Link Margin element or Link Measurement Request/Report frames.

\section{Conclusions}\label{Sec:Conclusions}
In this article, we highlighted different mechanisms incorporated into IEEE 802.11ay to support mmWave distribution networks. We introduced the main features of the use case, the associated design challenges, and the corresponding technical solutions adopted to meet the identified requirements. Specifically, we elaborated on the design of scheduling, beamforming, and link maintenance protocols.

Using unlicensed mmWave spectrum for fixed wireless access also has limitations. The imposed power restriction in unlicensed spectrum limits the transmission range of all nodes. The need of coexistence schemes to share the medium with incumbent devices in 60 GHz band is another concern. However, these limitations also present various research directions. For example, since mmWave communications are highly directional, there are more opportunities for spatial reuse, which allows simultaneous transmissions between different pairs of devices. Furthermore, the design of efficient scheduling algorithms that minimize interference within the network, and meanwhile preserve coexistence with legacy devices with random access, remains an open problem. Additionally, how to address the scalability of scheduling, beamforming, and link maintenance when the network grows large is also interesting to explore. Initial discussions about these future directions have emerged within IEEE 802.11 and will continue towards next generation mmWave standards design.

\ifCLASSOPTIONcaptionsoff
  \newpage
\fi

\nocite{*}
%\bibliography{References}{}

\begin{thebibliography}{1}
%\bibitem{11ad}
%IEEE 802.11 Working Group, ¡°IEEE 802.11ad, Amendment 3: Enhancements for Very High Throughput in the 60 GHz Band,¡± Dec. 2012.

%\bibitem{11ad-tutorial}
%T. Nitsche, C. Cordeiro, A. B. Flores, E. W. Knightly, E. Perahia, and J. C. Widmer, ¡°IEEE 802.11ad: Directional 60 GHz Communication for Multi-gigabit-per-second Wi-Fi," \emph{IEEE Communications Magazine}, vol. 12, no. 4, pp. 131-141, Dec. 2014.

\bibitem{11ay}
IEEE 802.11 Working Group, "Enhancements for Very High Throughput for Operation in License-exempt Bands Above 45 GHz," IEEE P802.11ay/D3.0, Feb. 2019.

\bibitem{11ay-tutorial}
Y. Ghasempour, C. R. C. M. da Silva, C. Cordeiro, and E. W. Knightly, "IEEE 802.11ay: Next-generation 60 GHz communication for 100 Gb/s Wi-Fi," \emph{IEEE Communications Magazine}, vol. 55, no. 12, pp. 186-192, Dec. 2017.

\bibitem{11ay-tutorial2}
P. Zhou et al. "IEEE 802.11ay based mmWave WLANs: Design Challenges and Solutions," \emph{IEEE Communications Surveys and Tutorials}, 2018.

\bibitem{11ay-ChannelModel}
A. Maltsev, A. Pudeyev, A. Lomayev, and I. Bolotin, "Channel modeling in the next generation mmWave Wi-Fi: IEEE 802.11 ay standard." \emph{22th European Wireless Conference}, pp. 1-8. VDE, 2016.

\bibitem{11ay-phy}
C.R.C.M. da Silva, J. Kosloff, C. Chen, A. Lomayev, and C. Cordeiro, "Analysis and Simulation of the IEEE 802.11 ay Single-Carrier PHY," \emph{2018 IEEE International Conference on Communications}, pp. 1-6, 2018.

\bibitem{11ay-bf}
C.R.C.M. da Silva, A. Lomayev, C. Chen, and C. Cordeiro, "Beamforming Training for IEEE 802.11 ay Millimeter Wave Systems," \emph{2018 Information Theory and Applications Workshop (ITA)}, pp. 1-9, 2018.

\bibitem{11ay-mimo}
M. Kim, T. Ropitault, S. Lee, and N. Golmie, "Efficient MU-MIMO Beamforming Protocol for IEEE 802.11ay WLANs," \emph{IEEE Communications Letters}, 2018.

%\bibitem{11ay-UseCase}
%IEEE 802.11 Task Group ay, ¡°IEEE 802.11 TGay Use Cases,¡± \emph{doc. IEEE 802.11-2015/0625r7}, Jul. 2017.

\bibitem{TDD-UseCase}
M. Grigat, S. Sawhney, D. Tujkovic, S. Krauss, C. Lange, and O. Bonnes, "mmWave Distribution Network Usage Model," \emph{doc. IEEE 802.11-17/1019r2} (Accessed on Jun. 20, 2019), Jul. 2017.

\bibitem{FWA-1}
C. Dehos, et al. "Millimeter-wave access and backhauling: the solution to the exponential data traffic increase in 5G mobile communication systems?" \emph{IEEE Communications Magazine} no. 9, pp 88-95, 2014.

\bibitem{FWA-2}
Z. Pi, J. Choi, and R. Heath, "Millimeter-wave gigabit broadband evolution toward 5G: Fixed access and backhaul." \emph{IEEE Communications Magazine} vol. 54, no. 4, pp. 138-144, 2016.

\bibitem{FWA-3}
C. Enjamio, E. Vilar, and F. Perez-Fontan, "Rain scatter interference in mm-wave broadband fixed wireless access networks caused by a 2-D dynamic rain environment." \emph{IEEE Transactions on Wireless Communications} vol. 6, no. 7, pp. 2497-2507, 2007.

\bibitem{FWA-4}
C. U. Bas et al. "Outdoor to indoor penetration loss at 28 GHz for fixed wireless access." \emph{2018 IEEE International Conference on Communications}, pp. 1-6, 2018.

\bibitem{FWA-5}
C. Saha, M. Afshang, and H. S. Dhillon. "Integrated mmwave access and backhaul in 5G: Bandwidth partitioning and downlink analysis." \emph{2018 IEEE International Conference on Communications}, pp. 1-6, 2018.

\bibitem{Terragraph}
Facebook, "Terragraph: Solving the Urban Bandwidth Challenge," available at https://terragraph.com (Accessed on Jun. 20, 2019).


\end{thebibliography}
\bibliographystyle{ieeetr}


\section*{Biographies}
\begin{IEEEbiographynophoto}{Cheng Chen}
is a Wireless Standards Research Engineer with the Next Generation and Standards Group at Intel Corporation. He has been an active contributor to IEEE 802.11ay, and was heavily involved in the definition of its MAC and beamforming specifications. Dr. Chen received the Ph.D. degree from Northwestern University, Evanston, IL.
\end{IEEEbiographynophoto}

\begin{IEEEbiographynophoto}{Oren Kedem}
is a Systems Engineer in Convergence, RF and Connectivity Group at Intel Corporation, Israel. He participated in developing Intel's IEEE 802.11ad product and has been an active contributor to IEEE 802.11ay MAC layer specification. Oren received the B.Sc and MBA degrees from the Hebrew University of Israel, Jerusalem.
\end{IEEEbiographynophoto}

\begin{IEEEbiographynophoto}{Claudio da Silva}
is a Systems Engineer with the Next Generation and Standards Group at Intel Corporation. He has been an active contributor to IEEE 802.11ay, and was heavily involved in the definition of its PHY and beamforming specifications. Before joining Intel, he spent time with Samsung and with Virginia Tech as an Assistant Professor.  Dr. da Silva received the Ph.D. degree from the University of California, San Diego.
\end{IEEEbiographynophoto}

\begin{IEEEbiographynophoto}{Carlos Cordeiro}
is a Senior Principal Engineer \& Senior Director with the Next Generation and Standards Group at Intel Corporation. He is responsible for Intel's Wi-Fi standards activities, including standardization for unlicensed millimeter frequencies. He is a member of the Wi-Fi Alliance Board of Directors and serves as its Technical Advisor. He is the technical editor to the IEEE 802.11ay standard and was the technical editor to the IEEE 802.11ad standard. Dr. Cordeiro received several awards including the prestigious Intel Inventor of the Year Award in 2016 and the 2017 IEEE Standards Medallion. He is the co-author of two textbooks on wireless published in 2006 and 2011, has published over 100 papers in the wireless area alone, and holds over 200 patents.
\end{IEEEbiographynophoto}
\vfill

%\newpage



%\vfill


\end{document}